\documentstyle[12pt,aasms4]{article}

\def\HA{{H$\alpha$ }}

\def\HST{\it HST }

\def\VEL{\:{\rm km\:s^{-1}}}

\slugcomment{}

\lefthead{Blair, Sankrit, Raymond, Long} \righthead{Cygnus Loop Distance}

\begin{document}

\title{The Distance to the Cygnus Loop from {\it Hubble Space Telescope} 
Imaging of the Primary Shock Front\altaffilmark{1}} 

\author{William P. Blair\altaffilmark{2},
Ravi Sankrit\altaffilmark{2},
John C. Raymond\altaffilmark{3},
Knox S. Long\altaffilmark{4}}

\altaffiltext{1}{Based on observations with the NASA/ESA 
{\em Hubble Space Telescope}, obtained at the Space Telescope Science Institute,
which is operated by the Association of Universities for Research
in Astronomy, Inc., under NASA contract NAS5-26555.}

\altaffiltext{2}{Department of Physics and Astronomy, 3400 N. Charles St.,
The Johns Hopkins University, Baltimore, MD 21218}

\altaffiltext{3}{Harvard-Smithsonian Center for Astrophysics, 
60 Garden St., Cambridge, MA 02138}

\altaffiltext{4}{Space Telescope Science Institute, 3700 San Martin
Drive, Baltimore, MD 21218}


\begin{abstract}

We present a {\it Hubble Space Telescope}/WFPC2 narrow-band H$\alpha$
image of a region on the northeastern limb of the Cygnus Loop supernova
remnant.  This location provides a detailed example of where the
primary blast wave first encounters the surrounding interstellar
medium.  The filament structure is seen in exquisite detail in this
image, which was obtained primarily as an EARLY ACQuisition image for a
follow-up spectroscopic program.  We compare the $\HST$ image to a
digitized version of the POSS-I red plate to measure the proper motion
of this filament.  By combining this value for the proper motion with
previous measurements of the shock velocity at this position we find
that the distance to the Cygnus Loop is 440 (+130, -100) pc,
considerably smaller than the canonical value of 770 pc.  We briefly
discuss the ramifications of this new distance estimate for our
understanding of this prototypical supernova remnant.

\end{abstract}


 

\section{Introduction}

The Cygnus Loop supernova remnant (SNR) is an extremely important
laboratory for studying many astrophysical phenomena related to shock
waves and their interaction with the interstellar medium (ISM).  Its
proximity, its large angular size, and relatively small foreground
extinction all help make it an important object for studies across the
entire electromagnetic spectrum.  Our thoughts and understanding about
the Cygnus Loop and what it represents have evolved dramatically over
the last several decades, culminating in the current picture, recently
summarized by \cite{lev97}, of a cavity explosion of a fairly massive
star.

Yet while our understanding of the Cygnus Loop has evolved
dramatically, there are other aspects of this important object that
remain more in the realm of folklore and seem to carry on from one
generation to the next.  One of these aspects is the distance to the
Cygnus Loop, which is an important and basic datum that affects nearly
every other aspect or interpretation of this object.  The oft-quoted
value of this parameter is 770 pc, attributed to \cite{min58}. He
performed a velocity ellipse analysis of 37 filaments and used the
proper motion measured by \cite{hub37} of 0\farcs03 yr$^{-1}$ for the
bright optical filaments to determine this value.  Except for
occasional ``heretical'' suggestions such as those of \cite{sak83} (1.4
kpc) and \cite{bra86} (460 pc), nearly all other researchers have
assumed Minkowski's value for the distance.

In this paper, we derive a new distance to the Cygnus Loop based on
$\HST$ observations of a single filament on the extreme northeastern
limb of the remnant. We obtain a proper motion for the filament by
comparing the $\HST$ data to a digitized version of the POSS-I red
plate, and use previous data on this filament's shock velocity to
constrain the distance.  We briefly discuss the ramifications of this
new distance estimate for previous studies of this important object.


\section{Observations and Data Reduction}

The filament we have observed in the Cygnus Loop has been studied a
number of times previously with other instruments and ground-based
telescopes (e.g. \cite{ray83}, \cite{fes85}, \cite{lon92} and
\cite{hes94}).  Located at RA = $20^h 56^m 2\fs7$ and Dec = 31\arcdeg
56\arcmin\ 39\farcs1 (J2000), it is on the extreme northeastern edge of
the Cygnus Loop, about 5\arcmin\ ahead of the bright radiative
filaments seen in this region.  This is a region of so-called
`nonradiative' shock front where the primary blast wave is encountering
partially neutral preshock material (cf. RBFG).  The X--ray
emission from the Cygnus Loop is bounded by these faint
Balmer-dominated shock fronts, as can be seen in the data presented by
HRB and \cite{lev97}.

The imaging data reported in this paper were obtained on 1997 Nov. 16
with the WFPC2 camera on the {\it Hubble Space Telescope}.  The image
was obtained primarily for use as an EARLY ACQ exposure, as part of our
Cycle 7 $\HST$ STIS campaign on this same filament.  However, we
expected the image to be interesting scientifically as well, and
devoted three orbits out of our program for this purpose.  We have
worked directly with the calibrated data extracted from the Guest
Observer tape provided by the STScI, using tasks available in the
IRAF/STSDAS environment\footnote{IRAF is distributed by the National
Optical Astronomy Observatories, which is operated by the Association
of Universities for Research in Astronomy, Inc. (AURA) under
cooperative agreement with the National Science Foundation.  The Space
Telescope Science Data Analysis System (STSDAS) is distributed by the
Space Telescope Science Institute.}.  The SNR filament was placed so
that it crossed the WF2 and WF3 CCDs.  We used two exposures per orbit
and the F656N filter, which is centered near the Balmer \HA line.  The
positioning was offset by $\Delta$x = $\Delta$y = 10 WFC pixels
(1\arcsec) between each orbit.  The two exposures from each orbit were
combined individually to remove most of the cosmic ray events.  Then
the first and third orbit data were shifted to align with the data from
the second orbit, and the three orbits of data combined into the final
image.  This produced an image clear of cosmic rays and camera hot
pixels, although some effects from the `gutter' between the WF2 and WF3
chips are still visible when the resulting data are displayed at high
contrast.  The total integration time was 7400 s.  Since stellar
contamination is not severe and the filament is known to emit primarily
in H$\alpha$, no other filters were used.

Figure \ref{HSTIMAGE} shows a 720 by 1484 pixel (72\farcs0 by
148\farcs4) region from the combined data.  In this image, north is
toward the upper right corner (position angle 30.24\arcdeg\ from
vertical) and east to the upper left, as indicated.  The brightest star
at lower left is $\HST$ GSID 0269203438 at V=13.3 and position RA =
$20^h 56^m 7\fs41$ and Dec = 31\arcdeg 55\arcmin\ 35\farcs70 (J2000).
The filament stretches across the image as a ribbon of light, with
variable intensity along both its length and width.  Here we see the
primary Cygnus Loop shock wave as a nearly edge-on sheet, gently
rolling along our line of sight as it encounters very slightly
differing preshock densities.  We see no hard kinks or twists in the
shock front that would be indicative of larger density contrast
features (or a cloud/intercloud type morphology {\it ala}
\cite{mck77}).  Rather, if appears that the brightness variations over
the observed region are dominated by line of sight effects, with
brighter regions corresponding to deeper columns and/or multiple shock
crossings along a given line of sight.  The crispest regions of edge-on
shock material are at or below our ability to resolve with the 0\farcs1
pixels of the WFPC2 Wide Field CCDs.  Since the \HA emission is
expected to be formed very close behind the shock front ($<10^{14}$ cm;
cf. RBFG), we are truly seeing a `snapshot' of the Cygnus Loop
shock front as it encounters the preshock medium.  Several exceedingly
faint filaments are seen in projection behind the primary shock (toward
the bottom in Figure \ref{HSTIMAGE}).  These filaments presumably arise
from other locations on the primary shock front seen in projection.
Their extreme faintness may be due to lower preshock densities or lower
neutral fractions in the preshock gas at those positions, or it may simply 
be that the path length through the emitting region is smaller.


\section{Analysis}

In contrast to \cite{min58}, we determine the distance to the Cygnus
Loop based on the properties of just a single filament.  To do this we
use the best value for the shock velocity at the observed position and
a measurement of the proper motion of the filament.  For measuring the
proper motion we compare our $\HST$ image with a digitized version of
the POSS-I red plate taken about 44 years earlier.  We discuss these
topics in the sections below.

\subsection{Constraints on the Shock Velocity}

The Balmer line emission from nonradiative shocks, such as the one we
are considering, comes from neutral atoms that pass through the shock
front and are collisionally excited by electrons before becoming
ionized (\cite{che78}; \cite{che80}).  In this zone immediately behind
the shock front, a significant fraction of the neutral hydrogen atoms
undergo charge exchange with the hot post-shock ions.  This results in
the Balmer lines having two distinct components - a narrow component
with a thermal width representative of the pre-shock temperature and a
broad component with a velocity spread representative of the post-shock
ion temperature.  Since the post-shock ion temperature depends on the
shock velocity, the width of the broad component of the \HA line is a
diagnostic for the shock velocity.

The translation of the width of the broad component of the \HA line to
an actual shock velocity depends upon the equilibration mechanism
between ions and electrons in the post-shock region.  The shock energy
thermalizes 3/4 the bulk velocity of the pre-shock particles so the
increase in the temperature of the ions is higher than the increase in
the temperature of the electrons by the ratio of their masses.  The
temperatures eventually come into equilibrium via Coulomb collisions.
However, if there is rapid equilibration between the ions and
electrons, for instance  via plasma turbulence within the shock front
(or some other mechanism), then the increase in ionic temperature is
lower than otherwise.  Therefore, a given width of the broad \HA line
implies a higher shock velocity for the case of rapid equilibration.

The two component line structure has been observed for the filament we
are discussing, but with somewhat discrepant results.  RBFG
used the Whipple 1.5 m telescope and echelle spectrograph and a
2\farcs5 by 7\farcs5 aperture oriented east-west across the central
portion of the filament.  They determined $\Delta v_{narrow}$ = 31
$\VEL$ and $\Delta v_{broad}$ = 167 $\VEL$, which implies that the
shock velocity is 170 $\VEL$ for the case of Coulomb equilibration and
210 $\VEL$ for rapid equilibration.  Later HRB used the Kitt Peak 4 m
telescope with a long slit (200$\arcsec$ by 1\farcs2) echelle oriented
nearly along the length of the filament (see HRB Figure 4).  The width 
of the narrow component in their spectrum agrees with the RBFG
value.  However, they found $\Delta v_{broad} = 130 \pm 15 ~ \VEL$, 
significantly lower than the RBFG value.  The inferred shock 
velocity is then 130 $\VEL$ for Coulomb equilibration and 165 $\VEL$ 
for rapid equilibration.

Geometric considerations can also affect the observed broad component
width.  Slight non-tangencies (especially both into and out of the
plane of the sky combined) would be expected to widen the measured
broad component width compared with truly edge-on.  Any such broadening
in the observed profiles for this filament must indeed be very
symmetrical, owing to the well-centered narrow \HA component in both
the RBFG and HRB data sets.  HRB estimated the extent of non-tangencies
to be $\sim$~6\arcdeg\ (plus and minus to keep the broad and narrow
components centered), a number that is consistent with the apparent
bumps and wiggles viewed {\it along} the filament in Figure 1. The
widening occurs as the sine of this angle, allowing of order 20\%
broadening from geometric effects (worst case).  Reconstructing the
RBFG and HRB slits onto the resolved image in Figure 1 shows a high
filling factor of very nearly edge-on shock material in the HRB slit
and a larger fraction of non-tangent shock material in the RBFG slit.
This is in the right direction to account for much of the observed
difference in broad component width, and perhaps indicates the RBFG
shock velocity estimates (e.g. 170 -- 205 $\VEL$) are on the high side.

A different set of diagnostics for the shock velocity is the strength
of lines from high ionization species arising further downstream from
the H$\alpha$ zone.  Since the ionization is due to collisions, the
highest ionization stage reached by any element depends on the
temperature of the post-shock gas which in turn depends on the shock
velocity.  Also these lines are formed further downstream where Coulomb
collisions have in any case had time to equilibrate the ion and
electron temperatures, so their strengths are not as sensitive to the
equilibration mechanism.  The filament under discussion here was
observed with the {\em Hopkins Ultraviolet Telescope} (\cite{lon92})
and its spectrum showed strong O~VI $\lambda\lambda$ 1032,1038 and N~V
$\lambda\lambda$ 1239,1243 emission.  By comparing the observed line
strengths with shock model calculations, \cite{lon92} found that the
spectrum could be best fit by shock models with velocities 175 to 185
$\VEL$.  Lower shock velocities could not produce the observed O~VI
emission and higher shock velocities resulted in an unacceptably high
ratio of O~VI to N~V emission.

As HRB and \cite{lon92} discuss, their observations can be reconciled
if either the shock front is rapidly decelerating, or if there is rapid
equilibration of ions and electrons in a 170 $\VEL$ shock front.  If
the shock is indeed decelerating, then the shock velocity appropriate
for the last 50 years needs to be used in calculating the distance to
the remnant.  If the deceleration is so rapid that the velocity changes
significantly over a period of 50 years, that also would need to be
accounted for.

Given the uncertainties discussed above, we adopt $v_{shock} = 170 \pm
20 ~ \VEL$  as a reasonable estimate for the relevant shock velocity, and
use this in the distance calculation below.

\subsection{A New Proper Motion Measurement}

We determine the proper motion of the filament by comparing our $\HST$
image with a digitized scan of the POSS-I red plate of the region.
This scan, kindly provided by the Catalogs and Surveys Branch at STScI,
was performed with 15 $\mu$m pixels, corresponding to 1\farcs0 per
pixel.  The POSS image was taken on 1953 July 14 and the $\HST$ image
on 1997 November 16, giving us a temporal separation of 16195 days
($1.40 \times 10^9$ s) between the two epochs.

We obtain the proper motion of the filament by measuring the
perpendicular distance between selected stars and the local shock front
in both POSS and $\HST$ images.  At the resolution of the POSS image,
the shock looks smooth.  In contrast, the $\HST$ image shows that the
shock front has very complicated substructure.  Therefore for our
comparison we have convolved the $\HST$ image with a Gaussian of FWHM =
5\farcs4, which corresponds to the point spread function determined for
stars in the POSS image.  In Figure \ref{PROPM_I} we show two locations where
intensity profiles were taken along cuts passing through the shock
front and a suitable star.  In each case, the leftmost panel shows the
POSS image, the middle panel shows the smoothed $\HST$ image and the
right panel shows the original $\HST$ image.  (The regions shown in the
POSS images have the same size as the regions in the $\HST$ images -
all are 74\arcsec\ $\times$ 72\arcsec\ although the alignments differ
by about 20\arcdeg).  The intensity profiles were taken along the
length of the boxes shown, and averaged over the width.

\begin{figure}
\plotfiddle{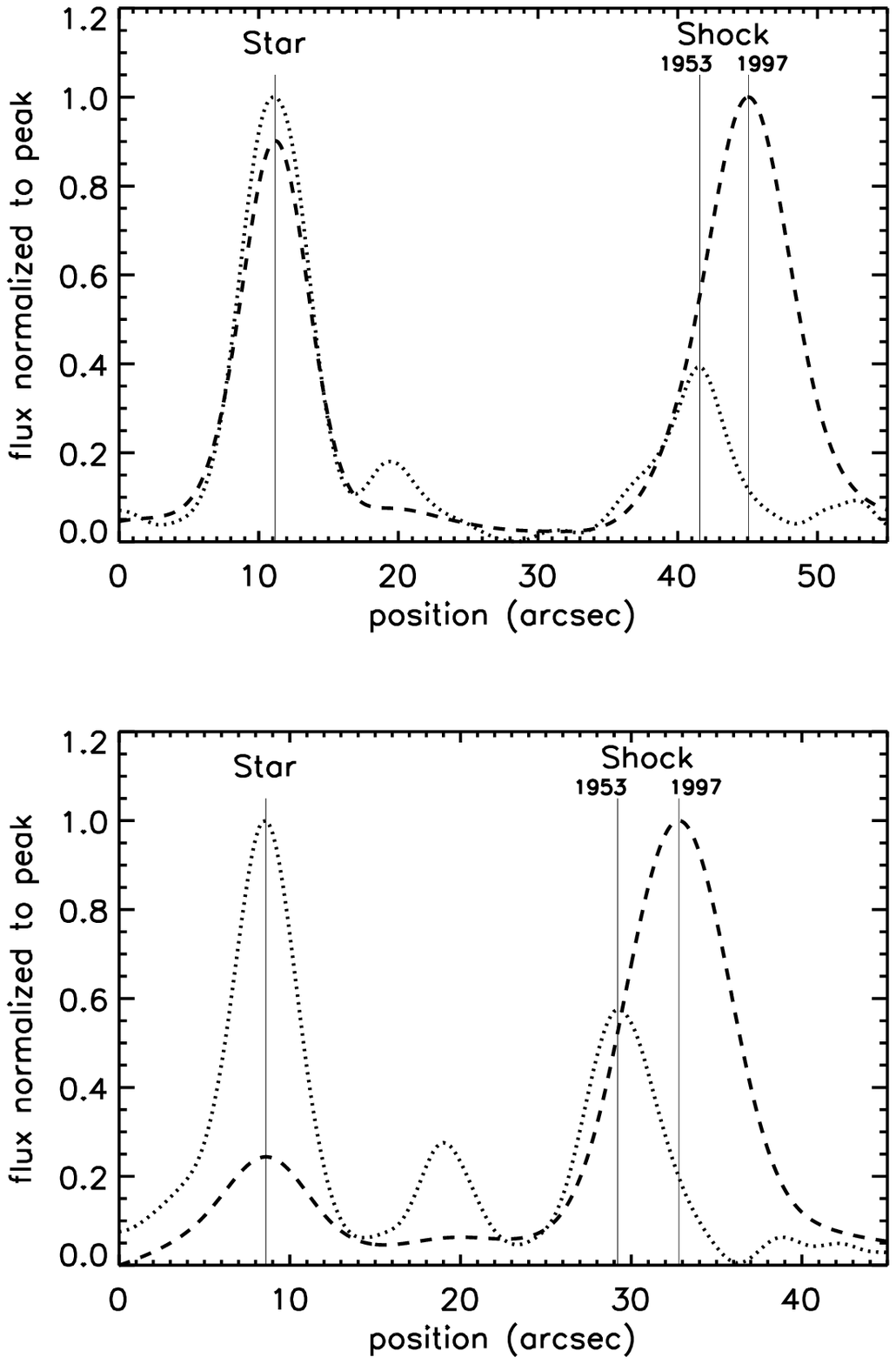}{7.5in}{0}{85}{85}{-254}{0}
\caption{\label{fig:fig1} Results of proper motion measurements.  The top 
plot is for Position 1 and bottom plot is for Position 2.  The measured 
proper motions are 3\farcs5 and 3\farcs6, respectively.
           \label{PROPM_P}}
\end{figure}

The results of our measurements are shown in Figure \ref{PROPM_P}.  For
each location, we have plotted the background subtracted, normalized
intensity profile along the cuts shown in Figure \ref{PROPM_I}.  The
dotted line shows the POSS profile and the dashed line the profile from
the smoothed $\HST$ image.  The star positions have been aligned, and
the advance of the shock front is clearly visible.  For Position 1 (top
panel), the shock front has advanced by 3\farcs5 and for Position 2
(bottom panel) by 3\farcs6.

Our use of stars as fiducials in measuring the proper motion of the
shock front is justified only if the stars themselves do not have a
high proper motion.  The best way to test for this effect would 
be to obtain astrometric solutions based on independently determined
positions of stars in the field.  Unfortunately, there is only one
catalogued star in the field of view of the HST image.  Therefore,
we have used the information in the respective FITS file headers to 
obtain astrometric solutions for both the digitized POSS and
HST images and compared the displacement of our fiducial stars relative
to a set of 14 other stars in the field.  We find that the nominal
changes in coordinates for our fiducial stars are not abnormal
compared with other field stars.  The standard deviation in the 
relative proper motion for all the stars is $\sim$ 0\farcs5.  For 
the specific stars used in our analysis, we find that the positional 
changes are 0\farcs5 and 0\farcs2 for the stars used at Position 1 and 
2, respectively.  The magnitude of errors thus introduced in the
measurement of shock proper motion is similar to those due to other
factors, as we discuss below.

Despite the poor resolution of the POSS image, this method should give
reasonably accurate results if the substructure of the shock has not
changed drastically between the two observations, and it is reassuring
that the results for two different locations give very similar values
for the proper motion.  To examine the effects of changes in the
filament substructure, we took a profile from the full resolution HST
image and changed the intensities of substructures within the shock
front in arbitrary ways and saw how that affected the location of the
peak in the smoothed profile.  We found that fairly extreme changes in
the substructure, such as completely eliminating the second strongest
peak in the Position 1 profile (ahead of the brightest band, see Figure
\ref{PROPM_I}), changed the derived proper motion by about 0\farcs5.
Another possible source of error is that we have taken profiles which
are perpendicular to an ``average'' shock front.  To estimate errors
caused by profiles not being normal to the local shock front, we
compared profiles at slightly different angles (passing through the
same star) and found that the derived proper motion could vary by about
0\farcs3.  Experiments with several other methods and crosscuts at
numerous other positions (using stars much farther from the local shock
front) all gave answers consistent with those given above, typically
within a few tenths of an arcsecond.  Hence, we adopt a value for the
filament proper motion of 3\farcs6 $\pm$ 0\farcs5 in 16195 days ($\sim$
44 years) for use below.

\subsection{Revised Distance to the Cygnus Loop}

The above velocity and proper motion can now be converted into a
distance, under the assumption that the motion of the filament is
directly transverse to the line-of-sight.  This assumption cannot be
far off for several reasons, including the appearance of the filament,
its position on the extreme limb of the SNR, and the fact that the
narrow Balmer line components in spectral data are well-centered on the
broad components (cf. HRB).  For our best estimate values of $v_{shock}
= ~ 170 ~ \VEL$ and proper motion = 3\farcs6, we find d = 442 pc.
Applying the uncertainties on these parameters as discussed above, we
derive an allowed range of 342 -- 573 pc (where the higher number
corresponds to the high velocity -- small proper motion limit and vice
versa).  We note that either a) substantial deceleration of this shock 
front over the time period of the measurement, b) widening of the 
observed broad Balmer component due to shock front geometry, or
c) some combination of both would only {\it lower} the appropriate 
velocity and hence {\it decrease} this distance estimate.  These 
uncertainties can clearly be reduced further, both by obtaining a 
second epoch of $\HST$ imaging data at 
some point and by better understanding of the electron -- ion 
equilibration and possible deceleration of the shock front.

It should be noted that the shock velocity estimates implicitly assume
that the kinetic energy dissipated in the shock front is transformed
into thermal energy of the ions and electrons.  If a large fraction of
the shock energy is used to accelerate cosmic rays, a higher shock
speed is required.  \cite{bou88} present a cosmic ray dominated shock
model for the filament in question with a shock speed of 365
km~s$^{-1}$.  However, the shock precursor in this model reaches far
too high a temperature to be consistent with the H$\alpha$ profile
(HRB), and it is likely that only $\sim$ 10\% of the shock energy goes
into non-thermal particles.  Thus consideration of cosmic ray
acceleration might increase the $v_{shock}$ estimate by 5\%.

While considerable uncertainty remains in the distance estimate, it is
clear that the canonical value of 770 pc is no longer tenable.  It is
of interest to note that \cite{bra86} concluded d = $460 ~ \pm ~ 160$
pc more than a decade ago, based on Hubble's (1937) and Minkowski's
(1958) original data but fitting for the best {\it mean} expansion
velocity instead of using the extreme of the velocity ellipse (as done
by Minkowski).  Also, \cite{shu91} estimated a distance of 600 pc,
but with a large uncertainty that extended upward and downward by a
factor of two.  Taken in this light, the distance derived here is not
out of line with the existing measurements for the bright filaments.

\section{Concluding Remarks}

A distance of $\sim$440 pc to the Cygnus Loop has some obvious and
important ramifications for the determination of the Cygnus Loop's
basic physical properties.  Quantities that depend linearly on distance
should be reduced by a factor of $\sim$0.6, while properties depending
on $\rm d^2$ will decrease by a factor of three.  At a distance of 440
pc, 1\arcsec = $0.6 \times\ 10^{16}$ cm and the angular dimensions of
the Cygnus Loop (2.8\arcdeg\ by 3.5\arcdeg; cf. \cite{lev97})
corresponds to linear dimensions of 21.5 pc by 27 pc.  Hence, the fact
that the crispest regions of edge-on shock in Figure 1 reduce to a
single WFC pixel or less places an upper limit of $0.6 \times\ 10^{15}$
cm on the size of the \HA emitting region behind the shock, still in
keeping with expectations (cf. RBFG).  Centered at galactic
latitude -8.6\arcdeg, a z distance of about 66 (d/440) pc is now
appropriate, placing the SNR much closer to the galactic mid-plane.
The inferred X--ray luminosity (\cite{ku84}) drops to $\rm L_{x}(0.1-4
keV)~=~3.6~\times~10^{34}~(d(pc)/440)~ergs~s^{-1}$.  Other parameters
can be similarly adjusted from the literature.

The smaller distance will also have ramifications for models such as
the ``cavity explosion'' picture put forward most recently by
\cite{lev97}.  Since the cavity does not have to be as large, the
inferred precursor star does not have to be as early as B0.  Also, a
smaller radius indicates a smaller age for the SNR would be
appropriate. \cite{ku84} determine a Sedov age of 18,000 yrs, which
reduces to 5000 (d(pc)/440) years.  Our point here is not to argue for
a Sedov model, but to simply point out that, as the ``prototypical
middle-aged SNR,'' perhaps the Cygnus Loop should be considered to be
on the young side of middle-aged.

It is interesting to note that another galactic SNR has recently
undergone a similar contraction in its distance estimate.  The Vela SNR
has been assumed to be at a distance of 500 pc for many years, based on
a crude estimate by \cite{mil68}. (Interestingly, this distance
estimate was at least partially based on the assumed `known' distance
to the Cygnus Loop!)  Several author's over the last decade have
provided reason to suspect a closer distance may be appropriate for
Vela, and a recent absorption line study to stars of known distance
toward Vela (Cha, Sembach, \& Danks 1999) solidifies this result:  Vela
appears to be at a distance of only 250 pc or so.

That the distances to two of the most intensely studied galactic SNRs
could be off by a factor of order two only serves to accentuate the
need to exercise caution when performing a comparative analysis of
galactic SNRs.  At their revised distances, the linear sizes of the
Cygnus Loop and Vela are quite similar, and yet their observed optical
and X-ray morphologies are quite different from one another.  Global
evolutionary studies for galactic SNRs will continue to be fraught with
uncertainty until distances to individual objects can be determined
with reasonable accuracy.

Obviously, a second epoch of $\HST$ imaging data on this filament in a
few years would provide a superior proper motion analysis and allow
this result to be refined.  Our nominal filament motion estimate of
3\farcs6 corresponds to an expected 0\farcs082 per year, or more
than three WFC pixels motion in four years.  Such as comparison would
also allow a direct assessment of any changes in relative brightness of
the shock front as a function of position and remove any effect of this
on the proper motion determination.

\acknowledgments

We wish to thank Brian McClean, Barry Lasker, and others in the
Catalogs and Surveys Branch at STScI for providing the digitized POSS-I
data.  We also thank the anonymous referee for a timely report with
useful suggestions.  This work has been supported by STScI grant
GO-07289.01-96A to the Johns Hopkins University.



\clearpage
\newpage

\figcaption{
$\HST$/WFPC2 \HA image of a nonradiative filament on the 
northeast limb of the Cygnus Loop. (Compare to Figure 1 of \protect\cite{lon92}
for a ground-based view.) The region shown is 1484 by 720 pixels (148\farcs4
by 72\farcs0); north is at upper right, east to upper left.  The shock wave is
moving upward in this figure, with the interior of the SNR toward the bottom.
            \label{HSTIMAGE}}

\figcaption{
Cuts along which intensity profiles have been taken.  Top set of images
is for Position 1 and bottom set of images is for Position 2.  In each
case left image is digitized POSS, middle is smoothed $\HST$ and right is 
full resolution $\HST$ showing sub-structure.  Note that for position 2,
the reference star is very faint in the smoothed $\HST$ image, but is 
clearly seen in the full resolution $\HST$ image.
           \label{PROPM_I}}

\clearpage


\end{document}